\documentstyle[epsf,12pt]{article}

\newcommand{\no}{\noindent}
 
 \newcommand{\dw}{density waves}
\newcommand{\beq}{\begin{equation}} \newcommand{\eeq}{\end{equation}}
\newcommand{\ve}{\breve} 
\newcommand{\mt}{$M_{\oplus}$} 
 \newcommand{\beqa}{\begin{eqnarray}}
\newcommand{\eeqa}{\end{eqnarray}} 
\newcommand{\exc}{eccentricity} 
\newcommand{\sem}{semi-major axis} \newcommand{\e}{$\, \,$}

\begin{document}

\noindent{\bf  \huge The  Origin  and Nature  of Neptune-like  Planets
Orbiting Close to Solar Type Stars}

\vspace*{1cm}

\noindent{\bf Adri\'an Brunini and Rodolfo G. Cionco\footnote
{ Also at Universidad Tecnol\'ogica Nacional-Facultad Regional San Nicol\'as, 
Col\'on 332, 2900, San Nicol\'as, Argentina.}}

\vspace*{1cm}

\noindent{\it Facultad de Ciencias  Astron\'omicas y Geof\'\i sicas de
la Universidad Nacional  de La Plata, Paseo del  Bosque s/n, La Plata,
Argentina.  Instituto de Astrof\'\i sica de La Plata (IALP), CONICET.}

\vspace*{3cm}
\noindent Number of pages: 27

\noindent Number of figures: 3

\noindent Number of tables: 1

\newpage

\noindent  Propposed running  head: Origin  of  Neptune-like extrasolar
planets

\vspace*{2cm}

\noindent   Correspondence    should   be   directed    to:   Adri\'an
Brunini. Facultad de Ciencias  Astron\'omicas. Paseo del Bosque s/n La
Plata    (1900)   Argentina.    Phone:   +54-221-4236593/4    -   Fax:
+54-221-423-6591. E-mail; abrunini@fcaglp.unlp.edu.ar

\newpage

\centerline{ABSTRACT}

\vspace*{1cm}

\noindent The sample of known  exoplanets is strongly biased to masses
larger than the ones of the giant gaseous planets of the solar system.
Recently,  the discovery  of  two extrasolar  planets of  considerably
lower  masses around the  nearby stars  GJ 436  and $\rho$  Cancri was
reported. They are like our  outermost icy giants, Uranus and Neptune,
but in contrast, these new planets are orbiting at only some hundredth
of the Earth-Sun  distance from their host stars,  raising several new
questions  about  their  origin  and  constitution.   Here  we  report
numerical simulations of planetary  accretion that show, for the first
time 
through N-body integrations
that  the formation of  compact systems of  Neptune-like planets
close  to  the  hosts  stars  could be a  common  by-product  of  planetary
formation.  
We found  a 
regime  of planetary  accretion, in  which
orbital migration accumulates protoplanets in a narrow region around the
inner edge of the nebula, where they collide each other giving rise to
Neptune-like  planets.
 Our results suggest that, if a protoplanetary solar environment is 
common in the galaxy, the discovery  of  a  vast
population of this sort of 'hot  cores' should be expected in the near
future.\\


\newpage

\section{Introduction}

The most plausible hypothesis regarding the formation of giant planets
is furnished by the core  accretion model (Mizuno, 1980), where a solid
core grows  by accretion of  planetesimals (Safronov, 1969).   When the
core  mass  increases above  a  certain  critical  value (10-15  Earth
masses),  a violent  accretion  of  nebular gas  onto  the solid  core
starts, leading  to the formation  of a gaseous giant  planet (Pollack
{\it et al}.,  1996). The final mass of the  gaseous envelope depends on
several factors, including the amount of gas present in the primordial
disk, and  the ability of  the solid core  to reach the  critical mass
before the dissipation of the surrounding nebula.

Two extrasolar planets of  low mass were recently discovered orbiting
around the nearby  stars GJ 436 (Butler {\it et  al.}, 2004) and $\rho$
Cancri (McArthur  {\it et al.}, 2004).  They are like  Neptune in mass,
but  these new  planets  are orbiting  at  only 0.02  AU  and 0.04 AU 
respectively from their host stars.

Standard disk models show that  at the distances where the new planets
have been  found (Butler  {\it et al.},  2004; McArthur {\it  et al.},
2004),  the temperature  is 600 -- 2000 K  (depending on  the stellar
type). 
Although at these distances, the existence of solid material cannot be 
ruled out, the temperature is  
too hot as to expect  a substantial amount of
it as to form  large solid cores there. 
More likely, giant
planets like these  ones form at much larger  distances from the star,
and   subsequently  they  migrate   inwards.   Theory   and  numerical
simulations  have  shown  that  growing  protoplanets  can  experience
orbital  migration,   traveling  very  far  from   their  birth  place
(Goldreich and Tremaine, 1980; Lin and Papaloizou, 1986; Ward 1986; Ward
1997a).  In fact,  planetary migration is a natural  consequence of the
tidal  interaction  with  the  nebula.  The  most  plausible  mode  of
migration  for protoplanets is  inward
in a quasi circular orbit
 (Ward,  1986; Ward, 1988; Ward,  1997a; Tanaka
{\it et al.}, 2002; Tanaka  and Ward, 2004).
  The tidal interaction with
the star or  the truncation of the disk  by the stellar magnetosphere,
prevent the planets  to fall onto the star.   Planetary migration have
been invoked  to explain  close stellar companions  (Lin {\it  et al.},
1996) of  large   mass  ('hot  Jupiters').
Ward (1997b), have also discussed in situ formation of ice giants planets
by migration of embryos to the inner protoplanetary disk limit.
Besides, Ida and Lin (2004) have delimited  conditions for which ice giants could form
in a variety of circumstellar disks.  Laughlin {\it et al.} ( 2004a),  
show that whereas ice giants form readily around stars of all masses, Jupiter-like 
planets are difficult to form around the smallest stars. 
The  new ice planets discovered support the findings of Laughlin {\it et al.} (2004a); indeed, 
GJ436 is a M dwarf type star, but in the $\rho$ Cancri system (which is a 'solar' G8V star) 
are coexisting the new 'Neptune' with three other Jupiter-like planets.
  However,  N-body simulations  of
planetary  formation  showing how  Neptune-like  planets  as the  ones
recently discovered could form, have not been yet carried out. In this
paper,  we  report  the   first  numerical  simulations  of  planetary
accretion,  showing that  the formation  of Neptune-like  planets very
near  the central  star could be a common  by product  of  planet formation, 
in a standard nebular environment.

\section{Simulations}

We  have performed  a series  of N-body  numerical simulations  of the
accretion  of solid  cores,  that  include, for  the  first time,  the
dynamical effects of the gaseous  environment on them, as predicted by
the more  recent results  of the theory  of planet-disk  interaction 
(Ward, 1997a; Tanaka {\it et al.}, 2002; Tanaka and Ward, 2004). In the set 
of simulations reported here, we have explored a model whose initial mass
 is of 
the order of the mass expected in the minimum-mass nebular model (NM)
(Hayashi {\it et al.}, 1985).

For the accretion of solid cores, we  follow the most plausible model 
which is based  on the concept  of 'oligarchic'  growth (Kokubo and Ida, 1998;
Kokubo and  Ida, 2000). In this  model, only  few  large objects 
growth at almost comparable  rates, separated by amounts determined by
their  masses   and  distances  from  the   star.   The  gravitational
scattering  with  the  largest  protoplanets dominates  the  dynamical
evolution of the background planetesimals, which for this reason cease
to growth.  However, the planetesimal disk, although representing only
a fraction  of the total  solid mass, may contribute  to the further 
growth of migrating protoplanets.
We have  performed four numerical  simulations where 100 small
protoplanets  of 0.5  Earth  masses were  initially  placed on  nearly
circular,  very low-inclination orbits  around a  star of  one  solar mass ($M_*$).
They  were spread  from 5  AU  to 15  AU, their  mutual separation were 
generated  at random, but following a $r^{-1.5}$ profile.  

It is very difficult to handle the amount of solid material required for 
a NM ($\sim  70 M_{\oplus}$) --note that we must integrate bodies spread over 
an extended disk--;
thus, we are compelled to make some  approximations. Oligarchyc growth models predict 
(at an initial time) a variety of embryo masses at different heliocentric distances, but  
we assume a uni-modal embryonic mass distribution. 
The number and mass  of the embryos  cited above, are plausible initial conditions
in an protoplanetary disk model of a few minimum-masses (see, Thommes {\it et al.}, 2003). 
Although the time-forming of such embryos can be very large, even in a several NM model,   
 if the random velocities of the embryos are sufficiently damped, the formation of big 
 bodies can be  expected (Thommes {\it et al.}, 2003). 
We make the hypothesis that the 
tidal interaction with a NM gaseous disk (not  considered in oligarchyc growth studies) can produce
 substantial velocity damping  and radial  
mobility of big embrios as to  strength the growth process (see e.g., Tanaka and Ida, 1999). 
Besides, during  the epoch of the formation of  
these embryos the nebular environment  could have been denser, helping to the growth 
process. Nevertheless, we think that detailed time and mass dependence
 of the initial conditions may not be important to a bottom-up set of simulations.

In addition, a swarm of residual planetesimals was included, occupying
the  same  space  than  the  protoplanets.  To  make  the  simulations
numerically tractable,  we used  200 planetesimals of  0.1 Earth
masses, distributed with a surface density proportional to $r^{-1.5}$.
Although the planetesimals interact  with the protoplanets, we ignored
the  self-gravity between  them (see e.g., Cionco and Brunini, 2002). 
A  gaseous disk  consistent  with the
minimum mass  solar nebula model (total mass,  density and  temperature
distribution, and  scale height) was considered (Hayashi  {\it et al.},
1995).  On  the planetary  embryos, the disk  acts dynamically  on the
orbital semimajor axes and eccentricities, as prescribed by the theory
of density waves.   The gaseous disk was assumed  truncated at 0.1 AU
and gradually  dissipates at a  constant rate, in  such a way  that it
completely   disappears   after   $10^7$   yr,  as   observations   of
circumstellar disks around young stars indicates (Beckwith and Sargent,
 1996). 
The gaseous disk  also affects the
orbits of the planetesimals through  aerodynamic gas drag (Adachi {\it et al.}, 1976), 
but as if their radii were of 1000 km.  

To carry  out these simulations, we have adapted our
N-Body  hybrid  code  (Brunini and  Melita, 2002;  Cionco and
Brunini, 2002) to include all these effects. 
 Our  model of planet-disk interaction includes migration of  type I. 
 
We didn't follow the 'classical' approach of summing the  
corresponding component torque  
over each active resonance (see e.g., Cionco and Brunini, 2002);
 here, we follow the most efficient strategy to be applied either in
 the leap-frog part of the hybrid integrator as in the Bulirsch--Stoer numerical 
 scheme used in it, in order to prescribe the orbital evolution predicted by 
 density waves theory. We correct the embryo velocities through a Stokes non--conservative force, yielding the same 
\sem  \e ($a$) and \exc \e ($e$) evolution predicted by \dw \e up to first order in $e$,  
applied in a consistent fashion, taking into account 
the disk properties at the different zones reached by  migrating bodies.         

This force, acting on a protoplanet of mass $M_p$ and  mean motion $\Omega$, produces an 
acceleration ($A_S$) and must be applied following the 
gas--planet relative velocity direction ($\ve{v}_{rel}$) as in the case of aerodynamic drag (Adachi {\it et al}., 1976):

\begin{equation}
\label{Stokes}
A_{S} = K \, M_p \, \rho \,  c_s^{\alpha} \ V_{rel},
\end{equation}

\no where $K$ is a parameter; $\rho$ is the volumetric gaseous density; 
$c_s$ is the sound speed in the medium (assumed as gas temperature dependent); 
$V_{rel}$ is the modulus of the relative gas--planet velocity. 
In what follows, we assume that 
all the distance--dependent quantities are evaluated at the \sem \e of the perturber.

Adachi  {\it et  al.} (1976),  have  found  expressions  for the  
 mean variation of the orbital elements 
over one orbital  period due to Eq. (\ref{Stokes}). Rewriting 
 $ \rho \propto   \Sigma/ h$, being 
$\Sigma $ the gaseous surface  density, and $h$ the isothermal vertical scale height
 defined by: $h \propto  c_s / \Omega$,  we can write  up to the first order in $e$:

\beq 
\label{daS}
\frac{1}{a} \left \langle \frac{da}{dt}  \right \rangle \propto   -K \, M_p \,
\Sigma \,  \Omega \,   c_s^{\alpha-1} \,  (h/a)^2, 
\eeq

\beq
\label{deS}
\frac{1}{e}   \left  \langle  \frac{de}{dt}   \right  \rangle  \propto 
  -K  \, M_p \, \Sigma  \, \Omega \,c_s^{\alpha-1}. 
\eeq

The perturbing  acceleration --Eq. (\ref{Stokes})--, must reproduce the same evolution of  \sem \e and \exc \e 
 estimated by \dw. This estimates are:
(see e.g., Ward, 1986;  Ward, 1988; Ward, 1997a; Tanaka {\it et al.}, 2002; Tanaka and Ward, 2004):

\beq
\label{dadw}
\frac{1}{a}  \frac {da}{dt} \propto  - \frac{M_p}{M_*} \, \left(\frac{\Sigma \, a^2}{M_*}\right) \,
\Omega \, (a/h)^2, 
\eeq

\beq
\label{dedw}
\frac{1}{e}\frac{de}{dt}    \propto - \frac{M_p}{M_*} \, \left(\frac{\Sigma \, a^2}{M_*}\right) \,
\Omega \, (a/h)^4. 
\eeq

\no  Expressions (\ref{daS})--(\ref{deS}) and 
(\ref{dadw})--(\ref{dedw}) are equivalent adopting $\alpha=-3$, and 
 $K \propto  G^2$, being $G=\Omega^2 \, a^3 / M_{*} $,  the gravitational 
constant. 

Finally, we adopted an appropriated constant in the definition of the 
parameter $K$ in order to 
match 'exactly' the \sem \e and \exc \e  evolution result predicted by 
Eqs.(\ref{dadw})--(\ref{dedw}) in our simulated
disk.

In  our simulations, a 1 $M_\oplus$ planet migrating at 5 AU, 
reach the inner disk limit in  $\sim 1 \times 10^6$ yr,   
and the time required for the same body, to reach a circular orbit  
(i.e., in attain $e\sim 5\times 10^{-5}$), is of $\sim 5 \times 10^4$ yr.

The inner edge of the disk is an uncertain parameter, nevertheless,  
the migration can be halted by gas depletion near 0.1 AU (see Ida and 
Lin, 2004, and references therein); thus, we assume that a protoplanet 
reaching the 'real' disk limit, can go on migrating until its outer 
Lindblad resonances get out of the disk by a factor of $\simeq 2$ respect to 
the physical disk limit.  Then, according to the frictional approach, 
when  a protoplanet  reaches 0.05 AU, it  is assumed  that 
the inward  migration stops.  

In  these simulations, if a planetary core reaches 10 Earth masses, it  start to
accrete   a   gaseous   envelope   following   the   core-instability
model (Pollack {\it et  al}., 1996).  
The gas accretion is prescribed adding an additional mass to  the embryo at each timestep, 
following the standard (J1) model of Pollack {\it et al.}, 1996. 

In the simulations when the numerical Bulirsch-Stoer part of the integrator is used (i.e., 
to integrate accurately embryo-embryo or embryo-planetesimal 
close encounters,  close stellar approaches, or the swarm of bodies reaching the inner disk limit), 
the time step is automatically adapted. We use a tolerance of $10^{-12}$.  
Other effects, as tidal interaction 
with the central star, quadrupole distortion between the planets, or 
relativistic effects, where not considered in these set of simulations
(see e.g., 
Trilling {\it et al.}, 1998, Kiseleva-Egletton and Bois, 2001).  

\section{Results}

The combined results  of the four simulations are shown in  Fig.  1.  It is
 evident  that  several  cores  (only  one of the  planets  accrete
 negligible amounts  of gas) reach the  inner edge of  the simulated disk, 
where their inward migration stops,
  and then, they survive with the 
 \sem \e reported.
  The migration of the cores is always of
 type I because their masses are not  large enough as to open a gap in
 the disk.  
 By  10 Myr, when the disk  is assumed completely dispersed,
 several  'hot cores'  resembling the  ones  found around  GJ 436  and
 $\rho$  Cancri   are  present. 
During   the  first  stage   of  the
 simulations, they growth by  the accretion of planetesimals, which is
 possible for two  reasons: first, the fast type  I orbital migration,
 which  is shown in  Fig.  2,  allows the  embryos to  be continuously
 immersed  in  zones not  previously  depleted  of planetesimals  
 \footnote{
 Note that if the disk is turbulent, the cores can attain a random migration 
 according to Laughlin {\it et al.} (2004b) not considered in our model; 
 nevertheless, these 
 fluctuations are largely contained in the random walk introduced by 
 the perturbations between embryos.} 
 and
 second, the aerodynamic drag  maintains the planetesimals in relatively
 low excited orbits, a  condition needed for efficient accretion.  But
 the most important  growth regime is due to  mutual accretion between
 massive embryos during the last  stages of the process.  This happens
 because a fraction  of them migrated quickly to  a narrow zone around
 the  inner  edge of  the  disk  favoring  mutual collisions  and  the
 subsequent formation of Neptune-like  planets. 
 
Table 1 shows the final 'hot cores' obteined in each simulation (only 
planets with $M_p >  10 M_{\oplus}$ are included) at $t=$ 10 Myr. 
The planets of the simulations Nr2 and Pr1 are 
remarkably similar to the ones  found around GJ 436:
21 \mt \e with $a=0.0278$,  (Butler {\it et  al.}, 2004);
 and $\rho$ Cancri: 17.7 \mt \e with $a$=0.04, (McArthur {\it et  al.}, 2004).
In  Fig. 3,  we have reproduced the growing  process of two planets 
in the above mentioned simulations. 
An interesting question arising from the simulations, is the amount of solid mass 
engulfed by the star. Although  it is in general negligible, simulation Pr2, 
 the only one with two final cores, disperses toward the star about 
one  core mass of solid material.  
 We believe
 that our  simulations are the  first showing this growth  regime, that
 operates on type I migrating protoplanets, in a standard nebular environment. 
     If  the  new
 discovered planets  formed in a  scenario as the one  described here,
 they should be of ice-rock composition with only a thin atmosphere.

\section{Conclusions}

Giant gaseous planets as massive  as Jupiter, or even Saturn, were not
found in this set of runs, but in all the four simulations, compact systems
of  two or  more Neptune-like  planets were  produced.   Although this
result points to the existence  of compact planetary systems formed by
small  giants, it  is in  contrast,  however, to  the system  orbiting
$\rho$  Cancri, which  already has  other three  Jupiter-like planets.
This fact could be due to the imposed restriction on the critical core
mass, for  fast accretion of gas,  which nevertheless is  not the only
possible scenario for the  core instability model (Stevenson, 1982). In
fact, according  to recent results of hydrodynamic  accretion of gas
around solid cores (D'Angelo {\it et al.}, 2002), planetary embryos may
accrete gas when the solid cores reach only 1-2 Earth masses.\\

Our  results  show that  the  formation  of  planets, and  systems  of
Neptune-like  planets around solar  type stars,  as the  ones recently
discovered, seems to be a natural  by-product of planetary formation, 
and encourage further studies statistically more robust, exploring the whole set of parameters i nvolved in the problem. 
Planetary migration  and the  value of  the critical  core mass  for substantial
accretion of gas, are  crucial factors in determining the distribution
of orbits and masses of the formed planetary systems.
Taking into account the standard scenario explored here, 
  we predict that a substantial population of 'hot  cores' with masses 
  between $\sim 14$ \mt \e and $\sim 24$ \mt \e placed at \sem $ < 0.05$ AU
    await to be discovered in the near future.
\vspace*{0.5cm}

\noindent{\bf Acknowledgements:} we acknowledge the finnancial support
by IALP, CONICET and AGNPCyT through grant PICT 03-11044 and the anonymous
referees that help us to improve the manuscript.

\vspace*{0.5cm} 

\noindent{\Large \bf References}

\vspace*{0.5cm}

\noindent Adachi, I., C. Hayashi and K. I. Nakazawa 1976.
 The gas drag effect on the elliptical motion of a solid body 
in the primordial solar nebula.	{\it Prog. Theor. Phys.}, {\bf 56}, 1756-1771. 

\noindent Beckwith,  S. V. W.  and A. I. Sargent  1996. Circumstellar
disks and the search for neighboring planetary systems.  {\it Nature}
{\bf 383}, 139-144.

\noindent  Brunini, A.  and M. D Melita 2002. The  Existence  of a
planet  beyond 50  AU and  the orbital  distribution of  the classical
Edgeworth-Kuiper-Belt objects.  {\it Icarus} {\bf 160}, 32-43.

\noindent  Butler,  R.  P.,  Vogt,   S.  S.,  Marcy,  G.  W.,  Fisher,
D. A. Jason,  T., Wright, G. W. H., Laughlin, G.  and J. J. Lissauer
2004. A Neptune-mass planet  orbiting the nearby M Dwarf  GJ 436. {\it
Astrophys. J.} (in press).

\noindent  Cionco,   R.  G.   and  A.  Brunini  2002. Orbital Migrations 
in planetesimal discs: $N$-body simulations and the resonant dynamical 
friction {\it  Mon.   Not.  Royal.   Astron.  Soc.} 334, 77-86

\noindent D'Angelo,  G.,  Kley,  W.  and T. Henning  2003.  Orbital
migration  and  mass accretion  of  protoplanets in  three-dimensional
global computations with nested  grids.  {\it Astrophys. J.} {\bf 586},
540-561.

\noindent Goldreich,  P.   and  S. Tremaine 1980.  Disk-satellite
interactions.  {\it Astrophys. J.} {\bf 241}, 425-441.

\noindent Hayashi, C., Nakazawa, K.  and Y. Nakagawa 1985. Formation of the
solar system. {\it Protostars and planets II}, 1100-1153.

\noindent Ida, S. and  D. N. C. Lin 2004. Towards a Deterministic Model of 
Planetary Formation I: a Desert in the Mass and Semi major axis Distribution 
of Extrasolar Planets {\it Astrophys. J.}{\bf 604}, 388-413

\noindent  Kiseleva-Eggleton L. and E. Bois  2001. Effects of perturbing forces on the 
orbital stability of planetary systems. {\it Astrophys. J.}
{\bf 553}, L73--L76.

\noindent Kokubo,  E.   and  S. Ida  1988.   Oligarchic  growth  of
protoplanets.  {\it Icarus} {\bf 131}, 171-178.

\noindent Kokubo, E. and S. Ida 2000. Formation of protoplanets from
planetesimals in the solar nebula.  {\it Icarus} {\bf 143}, 15-27.
 
\noindent Laughlin, G.,  Steinacker, A. and F. Adams 2004a. Type I Planetary Migration with MHD Turbulence
{\it Astrophys. J.}  {\bf 608}, 489-496

\noindent Laughlin, G., Bodenheimer P. and F. Adams  2004b. Core--Accretion Model 
Predicts Few Jovian--Mass Planets Orbiting  Red Dwarfs.
 {\it Astrophys. J.} {\bf 612}, L73-L76

\noindent  Lin,  D. N.  C.   and J. Papaloizou  1986.  On the  tidal
interaction between  protoplanets and the protoplanetary  disk.  III -
Orbital migration  of protoplanets.   {\it Astrophys.  J.}   {\bf 309}, 
846-857.

\noindent  Lin,   D.   N.    C.,  Bodenheimer,  P.    and  D. C. Richardson
 1996. Orbital migration  of the planetary companion of 51 Pegasi
to its present location. {\it Nature} {\bf 380}, 606-607.

\noindent McArthur,  B. E.   and 11 colleagues  2004. Detection  of a
Neptune-mass   planet  in   the   $\rho$  Cancri   system  using   the
Hobby-Eberley telescope. {\it Astrophys. J.} (in press.)

\noindent  Mizuno,  H. 1980.  Formation  of  the  giant planets.  {\it
Progress of Theoretical Physics} {\bf 64}, 544-557.

\noindent Pollack, J. B.,  Hubickyj, O., Bodenheimer, P. and J. J. Lissauer
 1996. Formation of the giant planets by concurrent accretion of
solids and gas.  {\it Icarus} {\bf 124}, 62-85.

\noindent Safronov, V.  S.  1969. {\it Evolution of the Protoplanetary
Cloud and the  Formation of the Earth and  the Planets}, Moscow. Nauka
Press.

\noindent Stevenson, D. J. 1982. Formation of the giant planets.  {\it
Plan. Space Sci.} {\bf 30}, 755-764.

\noindent Tanaka, H. and S. Ida 1999. Growth of a migrating protoplanet. 
{\it Icarus}{ {\bf 139}, 350-366.

\noindent     Tanaka,      H.,     Takeuchi,     T,      and  W.  R.
Ward   2002.  Three-dimensional interaction  between a  planet  and an
isothermal  gaseous disk.   I.   Corotation and  Lindblad torques  and
planet migration. {\it Astrophys. J.} {\bf 565}, 1257-1274.

\noindent  Tanaka,  H.   and  W. R.  Ward  2004.  Three-dimensional
interaction  between   a  planet  and  an   isothermal  gaseous  disk.
II. Eccentricity  waves and bending  waves {\it Astrophys.   J.}  {\bf
602}, 388-395.

\noindent  Thommes,  E.  W.,  Duncan,  M.  J.,  and  H.  F. Levison
2003. Oligarchic  growth  of giant  planets.  {\it  Icarus} {\bf  161}, 
431-455.

\noindent Trilling, D., Benz, W.,  Guillot, T., Lunine, J., Hubbard, W. 
and A. Burrows 1998. Orbital evolution and migration of giant planets: 
modeling extrasolar planets. {\it Astrophys. J.} {\bf 500}, 428-439

\noindent  Ward, W.  R.   1986. Density  waves in  the solar  nebula -
Differential Lindblad torque. {\it Icarus} {\bf 67}, 164-180.

\noindent  Ward,   W.  R.   1997a.  Protoplanet  migration   by  nebula
tides. {\it Icarus} {\bf 126}, 261-281.

\noindent  Ward,   W.  R.   1997b.  Survival of Planetary Systems
{\it Astrophys. J.} {\bf 482}, L211-L214.

\newpage 

\centerline{\bf \Large Figure captions:}

\begin{itemize}

\item{Figure 1} Final planetary masses of the embryos as a function of the mean distance to
the star. The different symbols represent the planets obtained in each simulation after 10 Myr.  In the four simulations, planets of 10-24
Earth masses were produced very close to central star.

\item{Figure 2} Evolution of the semimajor axis of one of the planetary cores
in our simulations.
The orbital migration obeys to the  type I migration, as the theory of
disk planet  interaction predicts.  The migration almost stops  when the core
reaches the inner edge of the simulated disk. At this location the core suffers strong dynamical interactions with other embryos that also have reached the inner edge of the disk. These  interactions produce the observed jumps  after 5.5 My at short semimajor axes.

\item{Figure  3} Mass accretion  of two  Neptune-like planets  vs. the
mean distance  to the  host star.  Although  the embryos  accrete some
background planetesimals (small jumps in mass at large semimajor axes), the accretion of substantial amounts of mass
is due to the mutual coagulation between embryos, represented by the
large steps in mass once they reach the inner edge of the disk. The presence of more than one embryo in a narrow zone around the inner edge of the disk produce the jumpy evolution in semimajor axis shown in  Fig. 2.

\end{itemize}

\newpage

\begin{table}
\label{tabla1}
\begin{center}
\caption{ Results of the four simulations at 10 Myr: name of the simulation; total mass,  
gaseous mass, \sem \e and  \exc \e of the cores obtained;  $M_{eng}$ is the mass  of solid material 
engulfed by the star.}
\begin{tabular} {lccccccc} \hline \hline  
       & $M_{core} [M_{\oplus}]$ & $M_{gas}[M_{\oplus}]$& $a$  [AU]            & $e$                  & $M_{eng} [M_{\oplus}]$  \\ \hline\hline   
Nr1    & 11.00                   & 0.000              & $4.015\times 10^{-2}$& $2.495\times 10^{-2}$&       0.0                       \\ \hline
Nr2    & 23.62                   & 0.018              & $3.530\times 10^{-2}$&$18.572\times 10^{-2}$&       0.0                       \\ \hline
Pr1    & 17.90                   & 0.000              & $3.840\times 10^{-2}$& $0.058\times 10^{-2}$&       0.0                       \\ \hline
Pr2    & 14.10                   & 0.000              & $3.305\times 10^{-2}$& $1.393\times 10^{-2}  $&       0.0                     \\   
       & 14.70                   & 0.000              & $4.340\times 10^{-2}$& $1.613\times 10^{-2}  $&       8.3                     \\ \hline
\hline
\end{tabular}
\end{center}
\end{table}

\newpage

\begin{figure}[t]
    \begin{center}
        \leavevmode
        \epsfxsize=10cm
        \epsfysize=15cm
        \epsffile{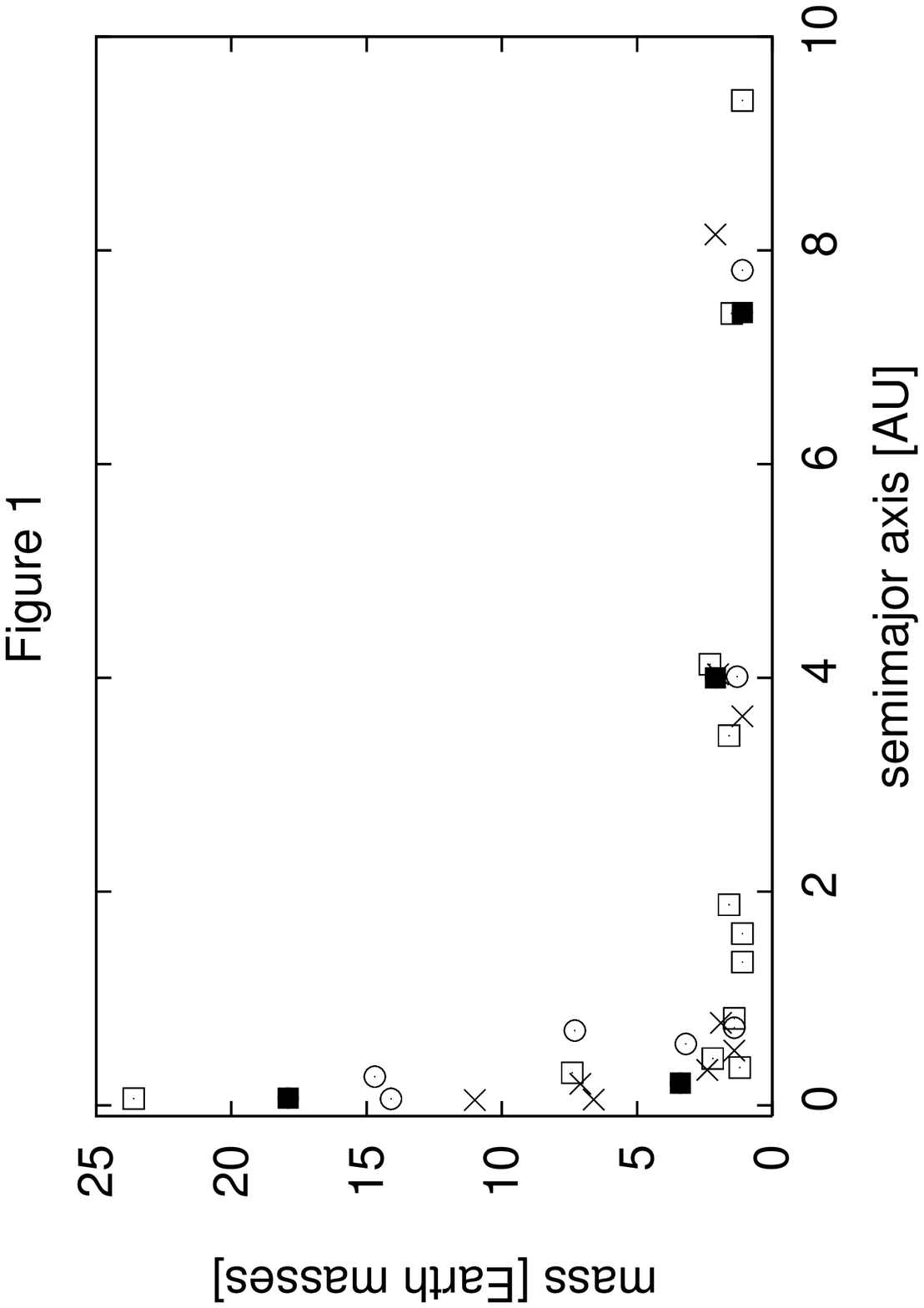}
     \end{center}
\end{figure}

\newpage

\begin{figure}[t]
    \begin{center}
        \leavevmode
        \epsfxsize=10cm
        \epsfysize=15cm
        \epsffile{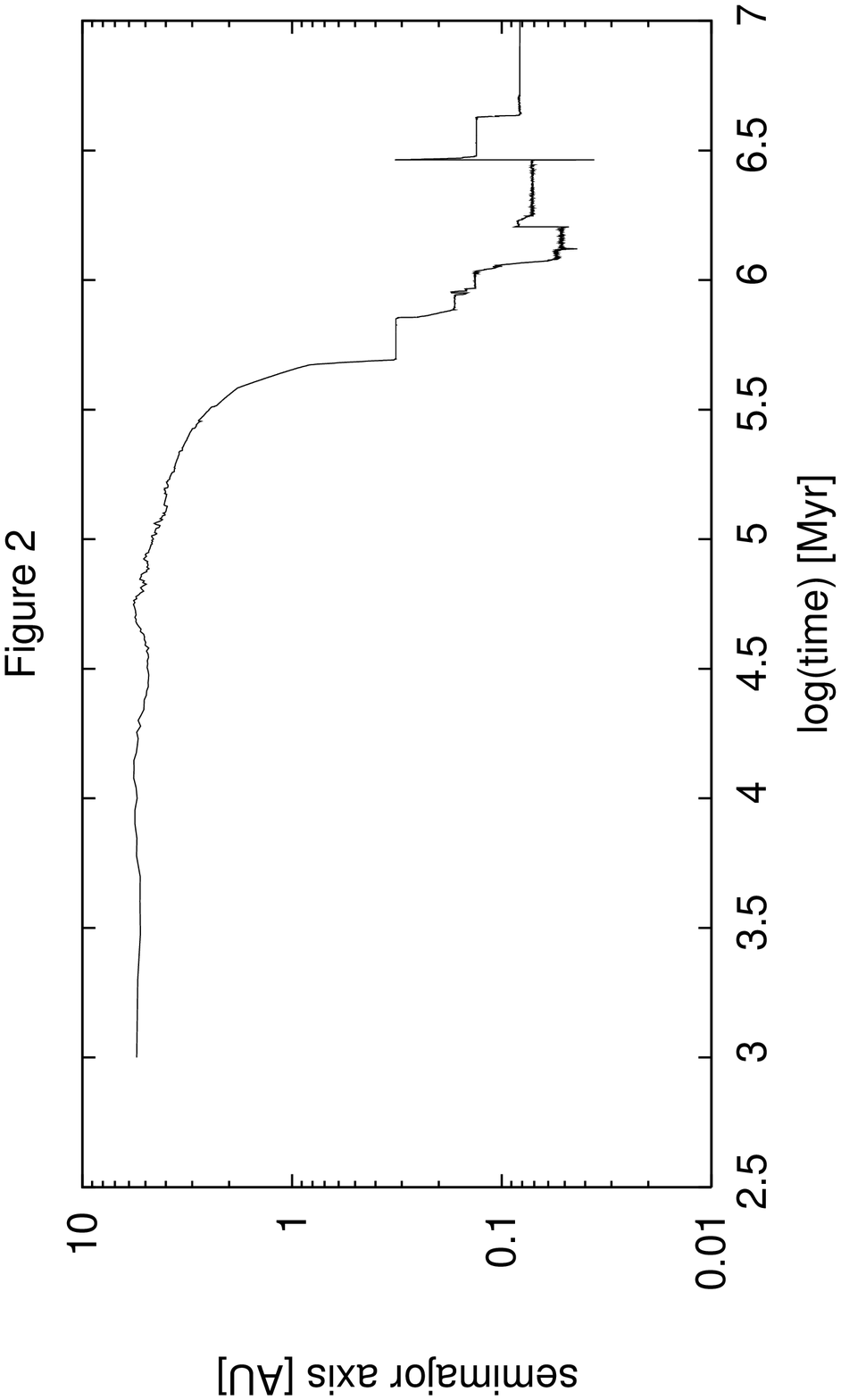}
     \end{center}
\end{figure}

\newpage

\begin{figure}[t]
    \begin{center}
        \leavevmode
        \epsfxsize=10cm
        \epsfysize=15cm
        \epsffile{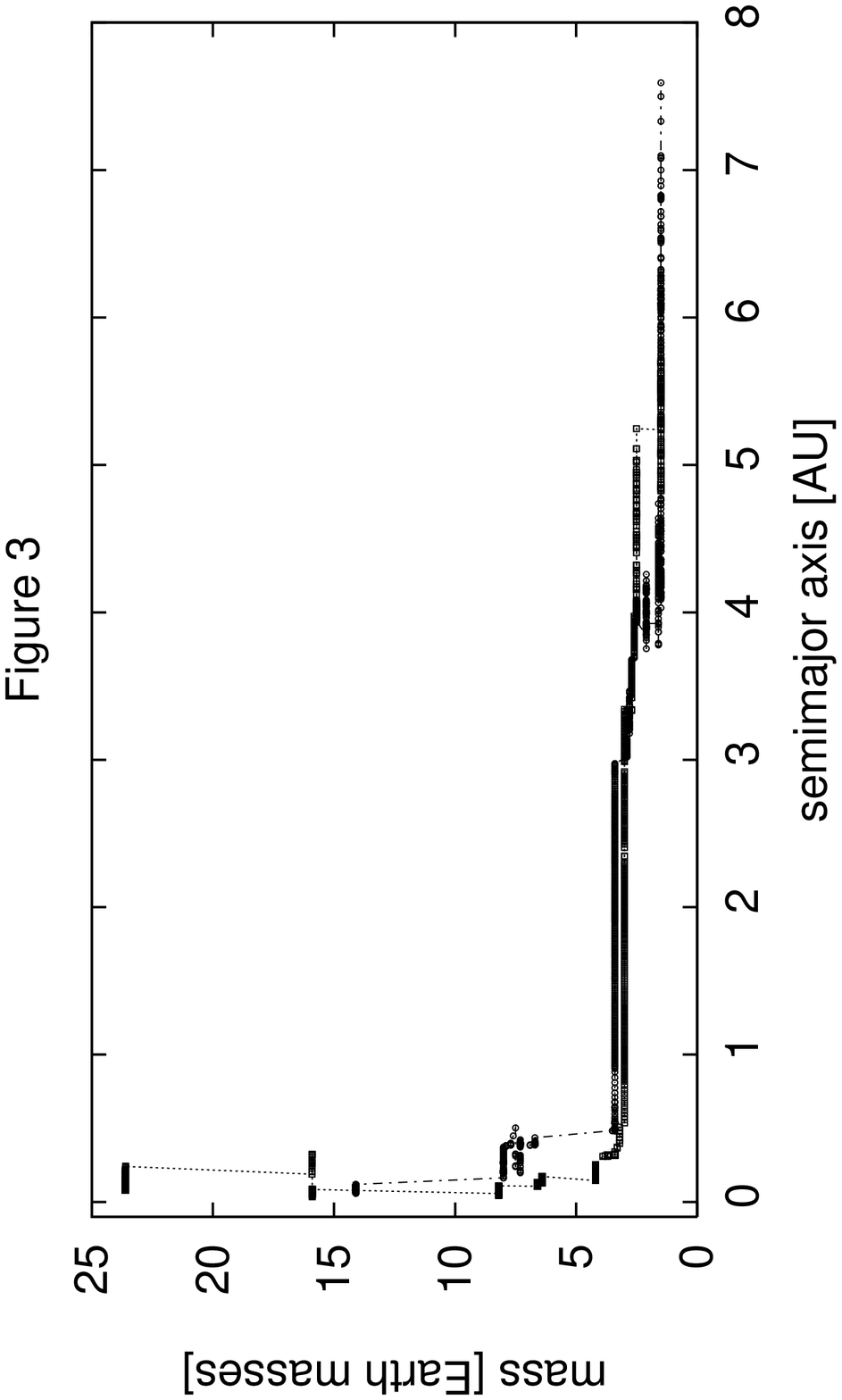}
     \end{center}
\end{figure}
\end{document}